\documentclass[aps,reprint,10pt,twocolumn,longbibliography,superscriptaddress]{revtex4-1}
\usepackage{xcolor}
 \pdfoutput=1
\usepackage{amsmath}
\usepackage{amssymb}
\usepackage{bbm}
\usepackage{graphicx}
\usepackage[countmax]{subfloat}
\usepackage{framed}
\usepackage{color}
\usepackage{hyperref}
\usepackage{epstopdf}
\usepackage{dsfont}
\usepackage{amsthm}
\usepackage{wrapfig}
\usepackage{relsize}
\usepackage{bm}
\usepackage{float}
\usepackage{enumitem}
\usepackage{soul}
\usepackage{subfig}

\newcommand{\red}[1]{{\color{red}}}

\usepackage{pifont}

\newcommand{\ba}{\begin{eqnarray}}
\newcommand{\ea}{\end{eqnarray}}

\begin{document}

\title{Diagnosing imperfections in quantum sensors via generalized Cram\'er-Rao bounds} 

\author{Valeria Cimini}\email{valeria.cimini@uniroma3.it}
\affiliation{Dipartimento di Scienze, Universit\`{a} degli Studi Roma Tre, Via della Vasca Navale 84, 00146, Rome, Italy}
\author{Marco G. Genoni}
\affiliation{Quantum Technology Lab, Dipartimento di Fisica ``Aldo Pontremoli'', Universit\`a degli Studi di Milano, 20133, Milan, Italy}
\author{Ilaria Gianani}
\affiliation{Dipartimento di Scienze, Universit\`{a} degli Studi Roma Tre, Via della Vasca Navale 84, 00146, Rome, Italy}
\affiliation{Dipartimento di Fisica, Sapienza Universit\`a di Roma, 00185, Rome, Italy}
\author{Nicol\`o Spagnolo}
\affiliation{Dipartimento di Fisica, Sapienza Universit\`a di Roma, 00185, Rome, Italy}
\author{Fabio Sciarrino}
\affiliation{Dipartimento di Fisica, Sapienza Universit\`a di Roma, 00185, Rome, Italy}
\author{Marco Barbieri}
\affiliation{Dipartimento di Scienze, Universit\`{a} degli Studi Roma Tre, Via della Vasca Navale 84, 00146, Rome, Italy}
\affiliation{Istituto Nazionale di Ottica - CNR, Largo Enrico Fermi 6, 50125, Florence, Italy}

\begin{abstract} 

Quantum metrology derives its capabilities from the careful employ of quantum resources for carrying out measurements. This advantage however relies on refined data post-processing, assessed based on the variance of the estimated parameter. When Bayesian techniques are adopted, more elements become available for assessing the quality of the estimation. Here we adopt generalized classical Cram\'er-Rao bounds for looking in detail into a phase estimation experiment performed with quantum light.
In particular we show that the third order absolute moment can give a superior capability in revealing biases in the estimation, compared to standard approaches.
Our studies point to identify a novel strategy that brings a possible advantage in monitoring the correct operation of high precision sensors. 
\end{abstract}

\maketitle

\section{Introduction}

The final goal of metrology is to infer the value of parameters that characterize a physical system with the best attainable accuracy. Usually the interest lies in the estimation of an interaction parameter associated with the unitary dynamics generated by a known physical process. A natural strategy to assess its value is to address the state of a probe after its interaction in the system. The information on the sought parameter is acquired by performing a series of repeated measurements ~\cite{Giovannetti1330, Giovannetti, Giov3}.

The key point of this estimation procedure is to efficiently extract all the available information from the measured data set, that is permeated by a random component. To this end, in classical parameter estimation, an estimator $\hat{\varphi}(\bold{x})$ is constructed in order to obtain a function of the data $\bold{x} = \{x_1, x_2,...,x_n\}$, that outputs the most accurate estimate of the investigated parameter $\varphi$ for a given set of data. 
 
In quantum parameter estimation ~\cite{doi:10.1142/S0219749909004839, PhysRevLett.96.010401}, the parameter is contained in the quantum state of a probe system which is then measured at the output using a suitable detection strategy \cite{review_natphot}.
In this scenario a further step is needed for maximizing the extracted information. The parameter is not the direct output of a measurement, but it needs to be inferred from a quantum observable. It is then essential to determine which observable carries the most information on $\varphi$. The problem of finding the optimal estimator will then be linked not only to the most accurate inference of the parameter value from the data, but also to the choice of the most suitable measurement scheme, over the class of all the possible \emph{positive operator valued measures} POVMs, as the one maximizing the precision~\cite{Helstrom, PhysRevLett.72.3439}. 

It becomes of paramount importance to identify upper bounds for the precision on the estimated parameter providing a figure of merit to compare strategies. This is introduced as the Fisher Information (FI), that represents the maximum amount of information, concerning the parameter of interest, that can be extracted from a measurement strategy. Its inverse bounds from below the achievable uncertainty, according to the Cram\'er-Rao bound (CRB) ~\cite{DEMKOWICZDOBRZANSKI2015345}, hence the optimal measurement is the one that maximizes the FI. A possible strategy to achieve the CRB is that of adopting a well-performing estimator such as a Bayesian one  \cite{Blandino,hradil95}. The goodness of the estimator used is then assessed by looking at its first and second order moment. The first gives the mean value and it is expected to deliver the true value of the parameter, the latter it is required to minimize the mean square error, bounded by the CRB, with respect to the true value. However, this stands true only if the estimator is unbiased. It can be insightful to inspect also the other order moments of the parameter probability distribution, with the aim of collecting further indications, detecting the presence of possible biases on the estimator.

Here we adopt the generalized CRBs, introduced by Barankin \cite{barankin1949} to assess quantum phase estimation. We investigate different moments of a Bayesian phase estimator, which is obtained using a polarization interferometer injected with N00N states. Our main result is that the third order absolute moment comes handy in detecting the presence of biases, which are unrevealed when using the standard approach. This work is organized as follows: in the next Section we describe the theoretical framework of estimation theory, and we introduce the generalized CRBs and the Bayesian estimator; in Section III we illustrate our results, both numerical and experimental; Section IV draws the conclusions. 

\section{Generalized Cram\'er-Rao bounds for detecting biased estimators}

In this section we formalize the problem of estimating an unknown parameter $\varphi \in \Phi$, via an indirect measurement of a different quantity $\mathcal{X}$. 
In practice one repeats the measurement $M$ times, obtaining a collection of measurement outcomes ${\bf x} =\{x_1,x_2,\dots,x_M\}$, that are assumed to be independent and identically distributed, drawn from the probability distribution $p(\bold{x}|\varphi)$. Consequently the likelihood of the whole experiment is given by

\begin{equation}
\mathcal{L}(\varphi) = p(\bold{x}|\varphi) = \prod_{i=1}^M p(x_i|\varphi).
\end{equation}

The estimator $\hat{\varphi}({\bf x})$ is defined as a map from the possible measurement outcomes ${\bf x}$ to the range $\Phi$ of possible values of the parameter $\varphi$. In particular, unbiasedness condition writes: 

\begin{equation}
\sum_{\bf x} p({\bf x }| \varphi) \left(\hat{\varphi}({\bf x}) - \varphi\right) = 0 \,,
\end{equation}

that is when on average it gives back the true value of the parameter. The variance of any unbiased estimator is proven to be bounded by the CRB \cite{cramer1946mathematical}
\begin{align}
\Big(\sum_x p(\bold{x}|\varphi)\big(\hat{\varphi}(\bold{x})-\varphi\big)^2 \Big) \ge \frac{1}{M \mathcal{F}[p({\bf x}|\varphi)]}\,, \label{eq:CRB}
\end{align}
where the Fisher information (FI) $\mathcal{F}[p({\bf x}|\varphi)]$ is defined as the second order moment of the log-likelihood function,
\begin{align}
\mathcal{F}[p({\bf x}|\varphi)]  =  \sum_x p(\bold{x}|\varphi)\Big(\frac{\partial}{\partial\varphi}\log p(\bold{x}|\varphi) \Big)^2 \,.
\end{align}
The definition of the FI can be extended to other (central) moments,
\begin{equation}
f_\alpha[p(\bold{x}|\varphi)] = \sum_x p(\bold{x}|\varphi)\Big|\frac{\partial}{\partial\varphi}\log p(\bold{x}|\varphi) \Big|^\alpha
\end{equation}
leading to the a generalized version of the CRB  in terms of the $\beta$-th absolute central moment~\cite{barankin1949, Li_2018,PhysRevA.97.022109}:
\begin{equation}
\Big(\sum_x p(\bold{x}|\varphi)\big|\hat{\varphi}(\bold{x})-\varphi\big|^\beta\Big)^{\frac{1}{\beta}}M^\frac{1}{2} \ge \frac{1}{\big(f_\alpha[p(\bold{x}|\varphi)]\big)^{\frac{1}{\alpha}}}.
\label{eq:gCRB}
\end{equation}
where $\frac{1}{\alpha} + \frac{1}{\beta} = 1$\,. As it is apparent the formulas above for $\beta=\alpha=2$, reduce to the standard definition of the FI, $\mathcal{F}[p({\bf x}|\varphi)]=f_2[p(\bold{x}|\varphi)]$, and the familiar CRB (\ref{eq:CRB}).\\
The CRB and its generalized versions are theorems that hold for unbiased estimators in the asymptotic regime of large $M$. As a consequence, if a violation of the bound is observed this may be a clear indication that these conditions are not met \cite{Brivio2010}. Specifically, either the collection of outcomes is too small, or the model employed for obtaining $p({\bf x}|\varphi)$ relies on wrong assumptions. This is the case when $p({\bf x}|\varphi)$ contains quantities subject to an incorrect pre-calibration \cite{Roccia:18}. The same strategy can be applied to analyzing the sensitivity of generalized CRBs to possible biases in the estimation, suggesting their usefulness as a diagnostic tool to check estimators' unbiasedness.\\
In particular we will focus on the Bayesian estimator \cite{olivares09, Teklu_2009}, that can be briefly introduced as follows. The estimated parameter is assumed to be a random variable distributed according to a \emph{prior} probability distribution $p(\varphi)$, representing the initial knowledge about its value. When a measurement is performed our information about the parameter changes and the (\emph{posterior}) conditional probability $p(\varphi|\bold{x})$ of the random variable $\varphi$, depending on the measurement outcomes $\bold{x}$, is updated. By using Bayes theorem one obtains
\begin{equation}
p(\varphi|\bold{x}) = \frac{p(\bold{x}|\varphi)p(\varphi)}{p(\bold{x})},
\end{equation}
where $p(\bold{x}|\varphi)$ is the likelihood function and $p(\bold{x})$ is the marginal probability of obtaining the data $\bold{x}$, that can be readily calculated by normalizing  the posterior distribution. We can then obtain the expected value of $\varphi$ (that for the \emph{ posterior} distribution is the Bayesian estimator) and the moments of the distribution as
\begin{align}
\hat{\varphi} &= \int_\Phi d\varphi \,\varphi \, p(\varphi|\bold{x})\,, \\ 
\Delta_{\beta}(\varphi) &= \int_\Phi d\varphi \big|\varphi-\hat{\varphi}\big|^\beta p(\varphi|\bold{x}).
\label{eq:BayesEstimator}
\end{align}
One can prove that the Bayesian estimator is asymptotically unbiased and optimal, that is for a large number of measurements $M$, the \emph{posterior} distribution becomes:
\begin{equation}
p(\varphi|\bold{x}) \simeq p(\varphi|M) = \frac{1}{N} p(\varphi) \prod_x p(\bold{x}|\varphi)^{Mp(\bold{x}|\varphi^*)},
\label{eq6}
\end{equation}
that is asymptotically approximated by a Gaussian \cite{braunstein92} with mean $\varphi^*$ and variance $\sigma^2 = 1/(M\mathcal{F}[p({\bf x} | \varphi^*)])$, where $\varphi^*$ is the true and unknown value of the parameter.

\section{Results}

\begin{figure}[H]
\includegraphics[width=\columnwidth]{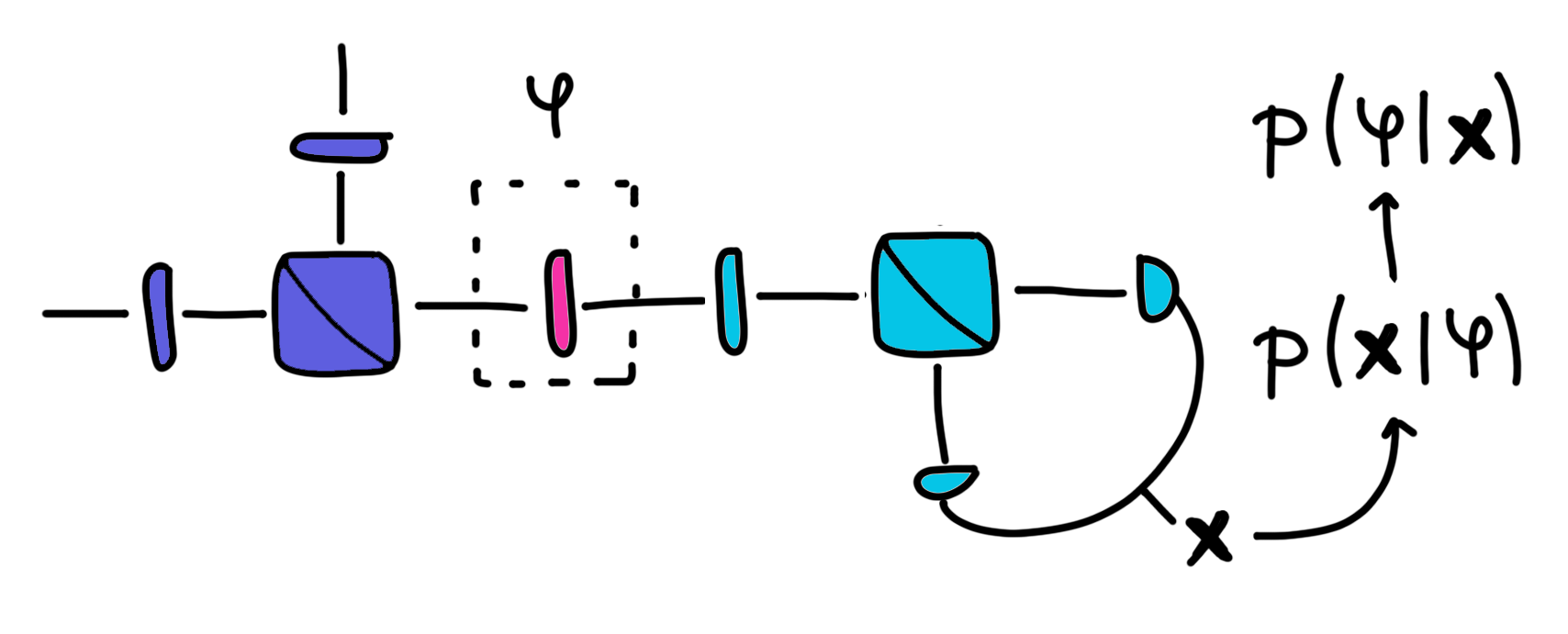}
\caption{Measurement scheme. A pair of photons is combined on a PBS in order to produce a N00N state in the circular polarization basis $\frac{1}{2}[(\hat a_R^\dag)^2+(\hat a_L^\dag)^2]|0\rangle$. The first HWP is used to impart the phase $\varphi$ between these two polarizations. The measurement apparatus performs the $4$-setting scheme \cite{Roccia:18,prlvale} allowing to reconstruct the \emph{posterior} probability.}   
\label{f:scheme}
\end{figure}

As a test-bed we consider a quantum phase estimation experiment \cite{hradil96,genoni12,dorner,Escher,Giov3,Holland,Mitchell,Higgins} carried out with a two-photon N00N state~\cite{PhysRevLett.107.219902,Nagata726,Afek879}. The measurement scheme is described in Fig. \ref{f:scheme}.
Starting with two photons with orthogonal polarizations, their combination on a polarizing beam splitter (PBS) leads to quantum interference, which can be observed on a rotated polarization basis. This shows up as an oscillatory behaviour of the coincidence counts at the outputs of a second PBS~\cite{PhysRevLett.59.2044}, that is observed when  a half-wave plate (HWP) is used to impart a phase $\varphi$. In our experiment, we set it to values in the interval $[0,180^\circ]$ in steps of $1^\circ$ ~\cite{prlvale}. 
The estimation is then performed collecting coincidence counts, in correspondence of each value of $\varphi$, using the four-measurement scheme in~\cite{Roccia:18,prlvale}. The data set consists in the collection of coincidence counts, relative to four different settings of the measurement HWP i.e. $\bold{\theta} = \{0, \frac{\pi}{16}, \frac{\pi}{8}, \frac{3\pi}{16}\}$.

The conditional probability to detect a coincidence event, that is to obtain the measurement outcome $\theta$ given the value of the phase $\varphi$,  is given by 
\begin{align}
p(\bold{\theta}|\varphi) = \frac{1}{4} \big(1+v_0\cos(8\theta-2\varphi)\big).
\label{lasolita}
\end{align} 
This conditional probability evidently relies on the pre-calibration of the visibility $v_0$ of the interferometer, and, as extensively explained in~\cite{Roccia:18}, an incorrect determination of the pre-calibrated visibility can affect the value of the phase parameter, resulting in a biased estimation~\cite{doi:10.1080/23746149.2016.1230476}.

In the following we will monitor the behaviour of the Bayesian estimator with respect to the generalized CRBs (Eq.~\ref{eq:gCRB}) of different orders  for $\beta = \{\frac{3}{2}, 3, 4\}$, to establish whether they are more affected by the visibility bias, as compared to the standard CRB ($\beta = 2$). We will first present some numerical simulations and finally present some experimental data.
\subsection{Numerical Simulations}

We perform numerical simulations of $N_{\sf exp}=400$ experiments, with each experiment corresponding to $M=2000$ measurements. We fix the interferometer visibility to $v_{\sf true}=0.95$, and we study the generalized CRBs for three different values of the phase to be estimated $\phi=\{ \pi/8 , 3\pi/16, \pi/4\}$. The phases have been chosen in order to consider their value where the FI is maximum ($\varphi=\pi/8$), minimum ($\varphi=\pi/4$), and an intermediate value ($\varphi=3\pi/16$). For each experiment, the $M$ outcomes are employed to construct the Bayesian estimator. We use a flat prior distribution $p(\varphi)$ that sets the limits of the integration region $\Phi=[0,\pi/2]$, and then we assume to have pre-estimated the interferometer visibility, and in particular we will consider different values $v_{\sf est}$ in the interval $[0.9,1]$.

\begin{figure}[H]
{\includegraphics[width=.9\columnwidth]{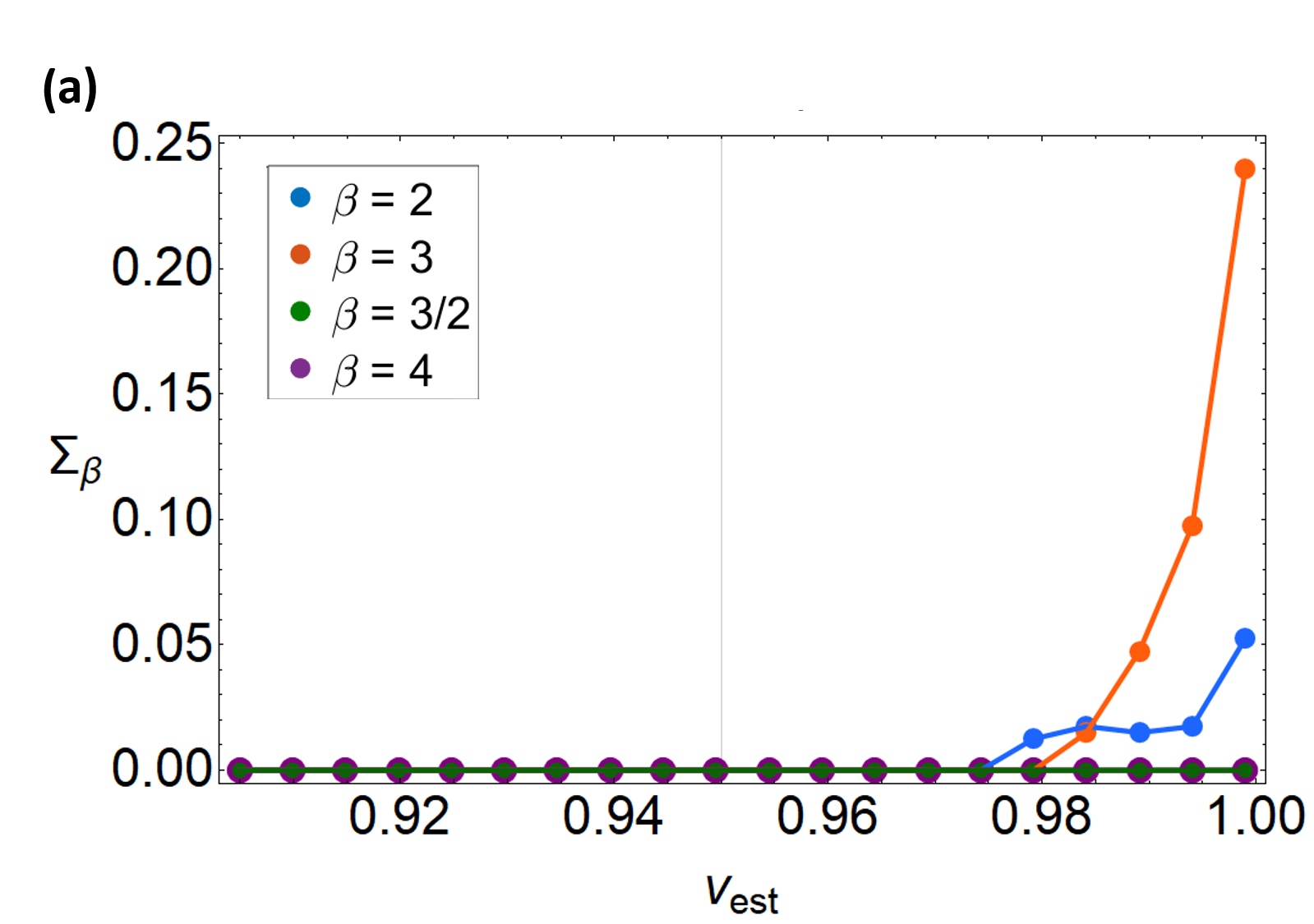}} \\
\vspace{0.2cm}
{\includegraphics[width=.9\columnwidth]{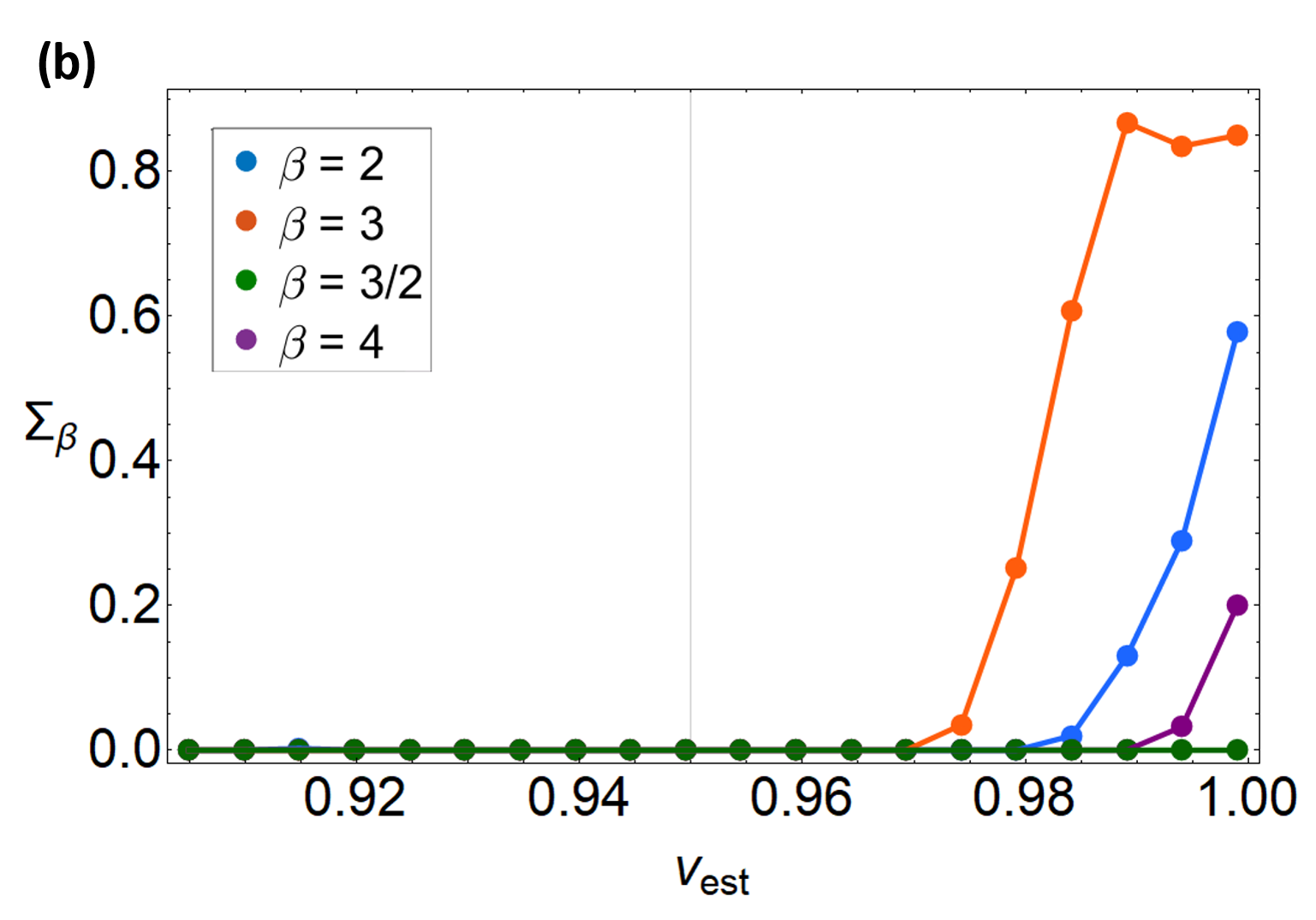}} \\
\vspace{0.2cm}
{\includegraphics[width=.9\columnwidth]{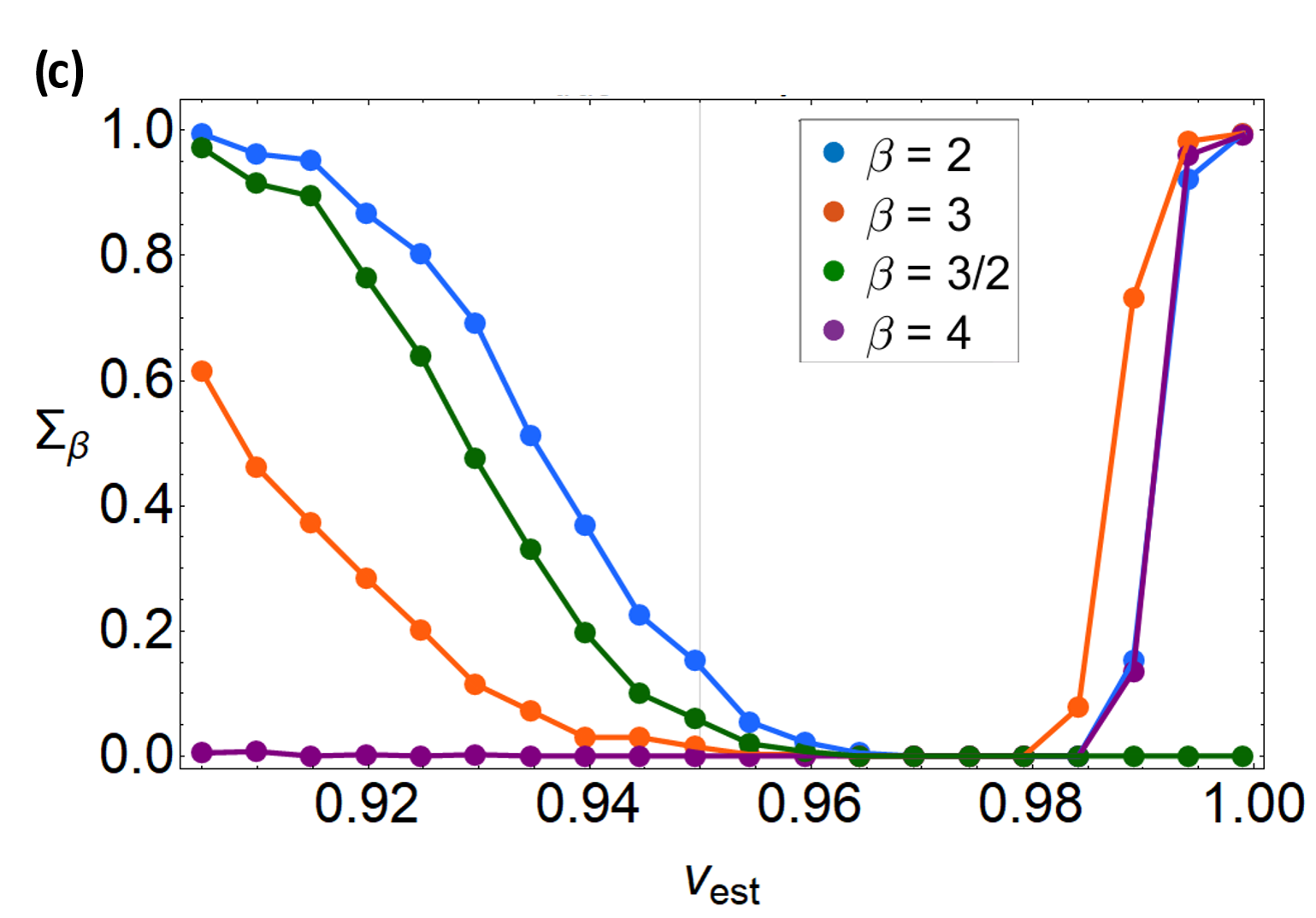}}
\caption{$\Sigma_\beta$ as a function of the pre-estimated value of the interferometer visibility $v_{\sf est}$, and for different values of the moment order. We considered different phases respectively in panel (a) $\varphi=\pi/8$, (b) $\varphi=3\pi/16$, and in (c) $\varphi=\pi/4 $, while the vertical line in each plot corresponds to the true value of the visibility $v_{\sf true}=0.95$.}
\label{f:numerics}
\end{figure}

By using Eqs. (\ref{eq:BayesEstimator}), we can calculate the central moments of the Bayesian estimator and verify whether the generalized CRBs are violated. We label as $\Sigma_\beta$ the fraction of experiments where the generalized CRBs in Eq. (\ref{eq:gCRB}) are violated. In Fig. \ref{f:numerics} we plot $\Sigma_\beta$ as a function of the estimated visibility $v_{\sf est}$, for different values of $\beta$ and for the three values of the phase $\phi$.   We will consider the bound to be violated outside the range $\pm 3\sigma_\beta$, where $\sigma_\beta$ corresponds to uncertainties for the moments $\Delta_\beta$, assuming a Gaussian form for the Bayesian posterior.

We first observe that in general the violations are more likely to happen for values of the phase where the FI is minimum. Remarkably, we also observe that generalized CRBs for $\beta\neq 2$ can be more efficient in detecting biases, respect to the standard CRB for particular values of $\varphi$ and $v_{\sf est}$. In particular this holds for $v_{\sf est} \geq 0.98$: while $\beta=3$ seems to be more efficient in this regime for each phase considered, we observe that for $\varphi=\pi/4$, the moment of order $\beta=4$ is as efficient as the standard $\beta=2$. Notice that in the case of $v_{\sf est} > v_{\sf true}$, it is more likely to obtain a biased estimation due to the functional form of the conditional probability $p(\theta | \varphi)$ employed to construct the Bayesian estimator: in this case, the experimental frequencies may achieve values unattainable with those expressed by Eq.(\ref{lasolita}).

Finally, for $\varphi=\pi/4$, that is in the regime where the Fisher information is minimal, we observe a non negligible fraction of violated CRBs, for $\beta=2$ and $\beta=3$, also when one considers the correct estimated visibility $v_{\sf est} = v_{\sf true} = 0.95$. This may be due to the fact that, in this regime, the Bayesian estimator has not reached its asymptotical optimality yet, and thus one would need a larger number of measurements $M$ to reach unbiasedness and to attain the CRB.

\subsection{Experiment}
A measurement of the visibility of the setup gives $v = 0.985 \pm 0.003$ at the beginning of the phase measurement and, as the experiment proceeds, it decreases to $v = 0.954 \pm 0.004$, due to setup misalignment over time. For each phase $\varphi$ we compute the four $\beta$ moments of the distribution ($\Delta_\beta$), for different  visibilities $v_{\sf est}$ in the interval between $[0.90,1]$, comparing their values to those corresponding to the generalized CRBs. 

\begin{figure}[H]
\begin{center}
{\includegraphics[width=.45\columnwidth]{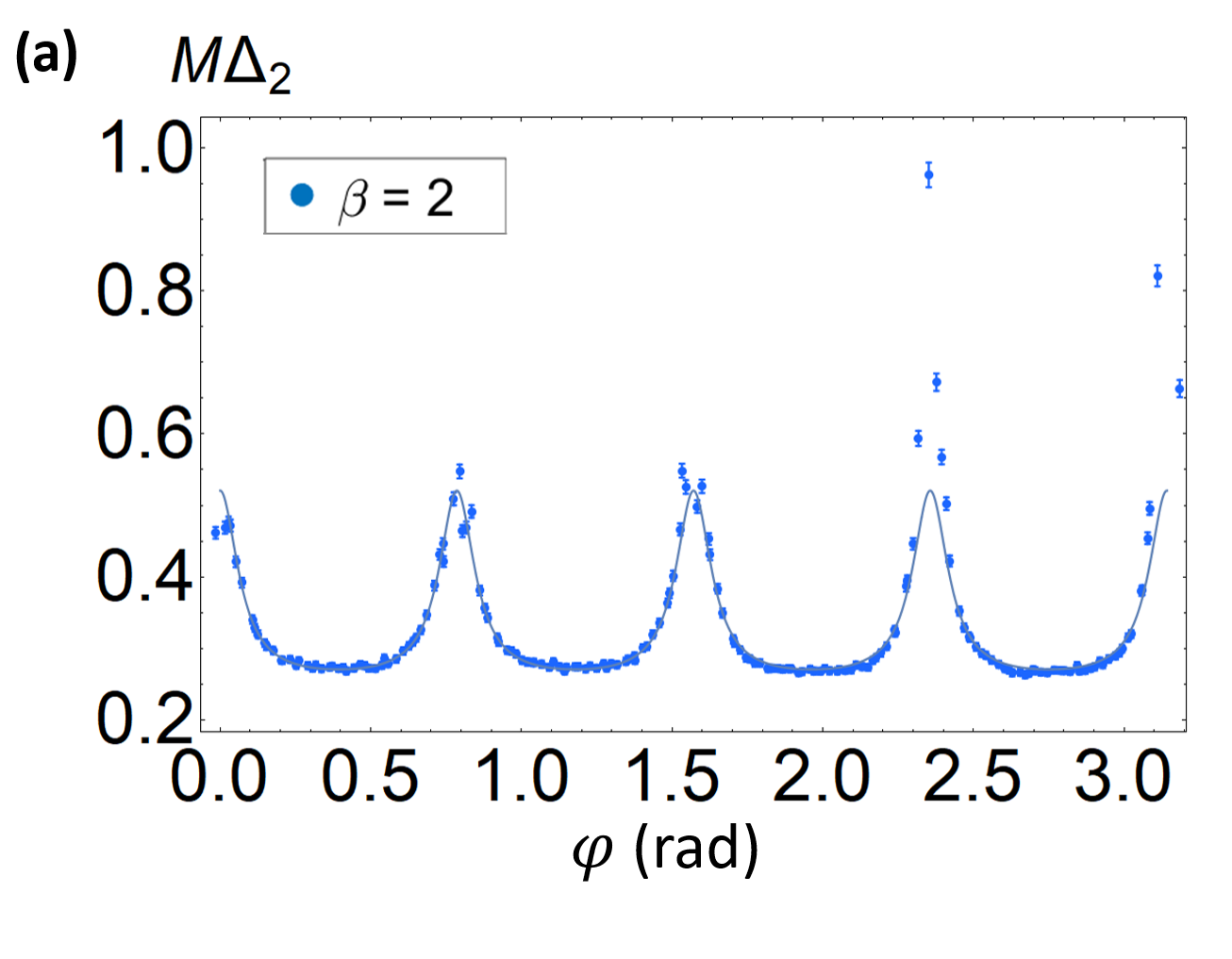}} \quad
{\includegraphics[width=.45\columnwidth]{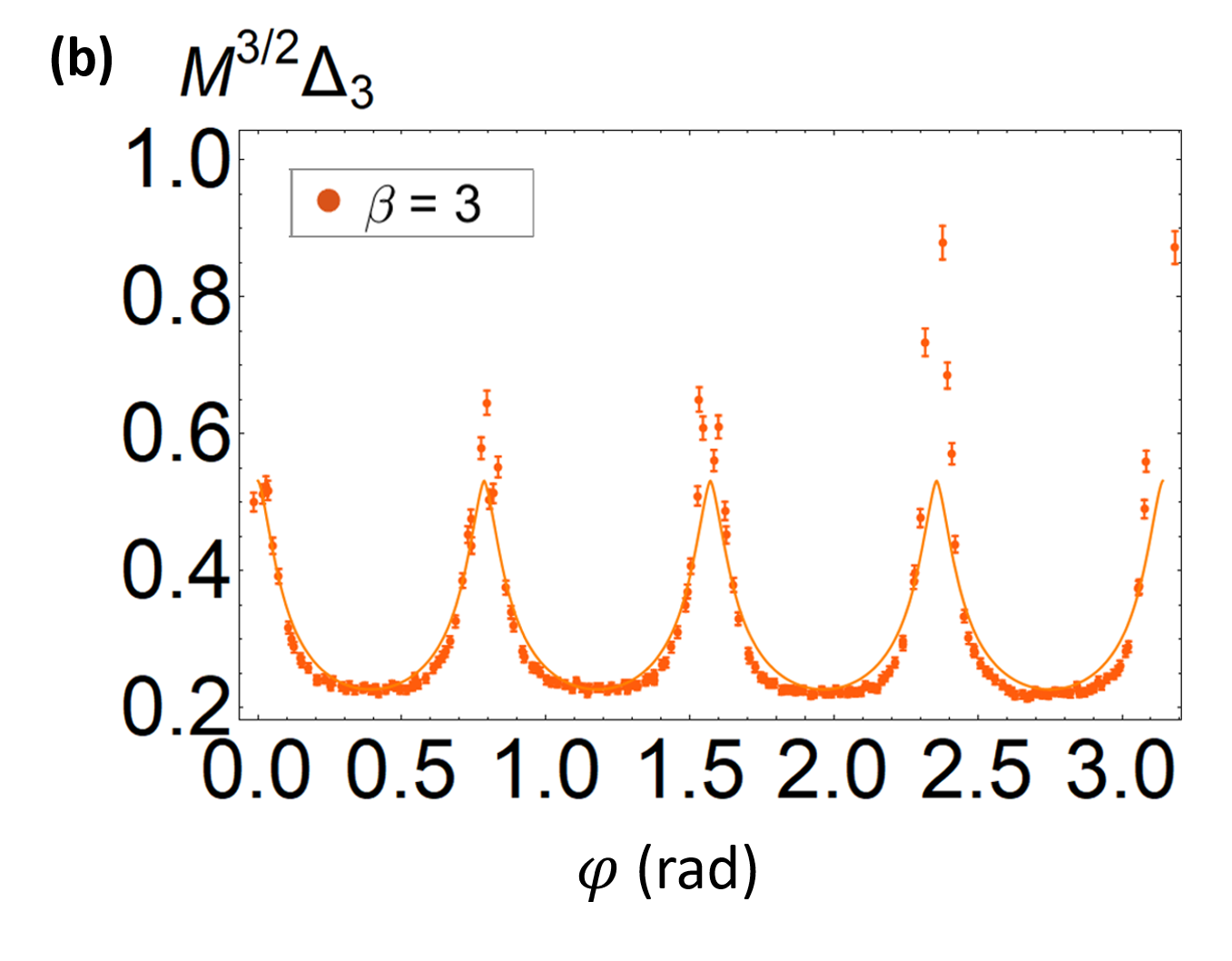}} \\
{\includegraphics[width=.45\columnwidth]{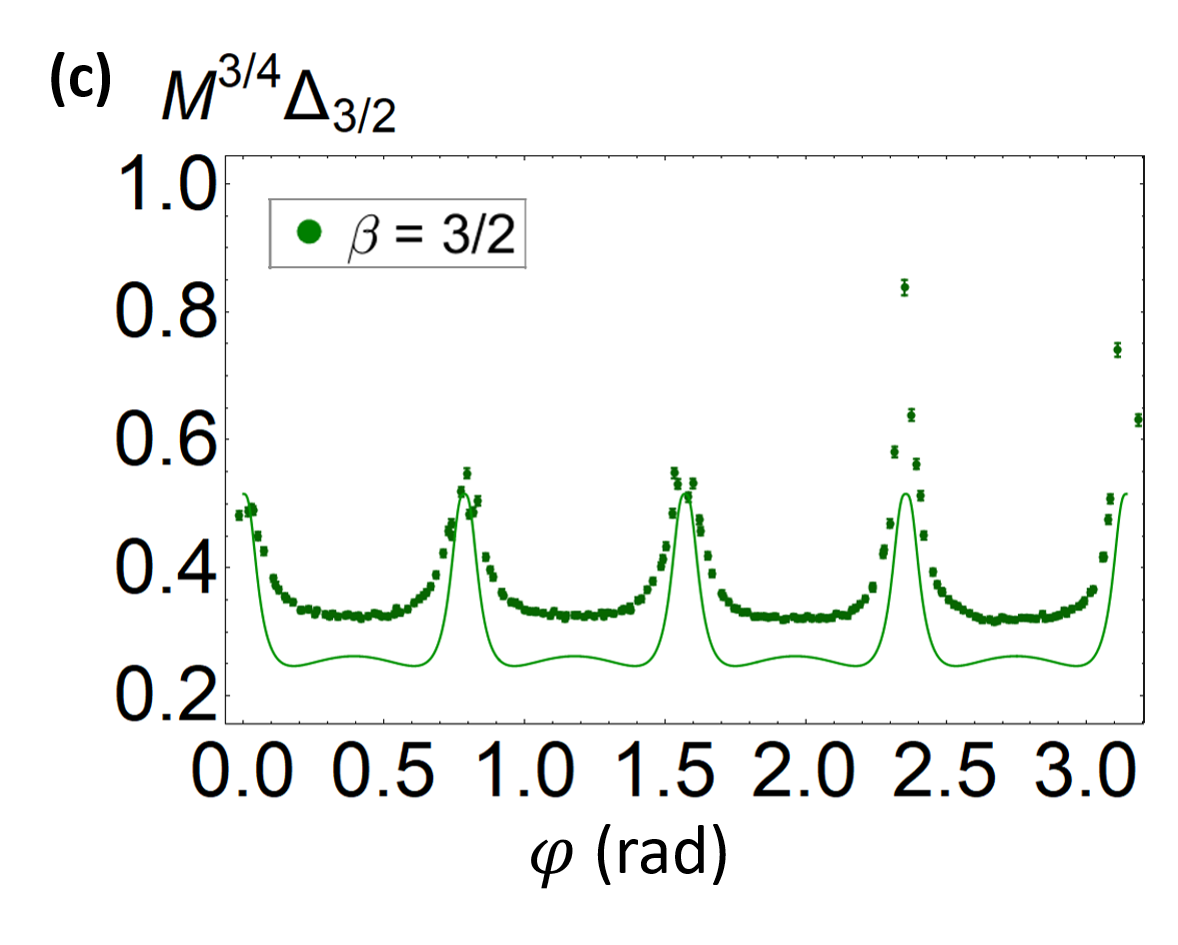}} \quad
{\includegraphics[width=.45\columnwidth]{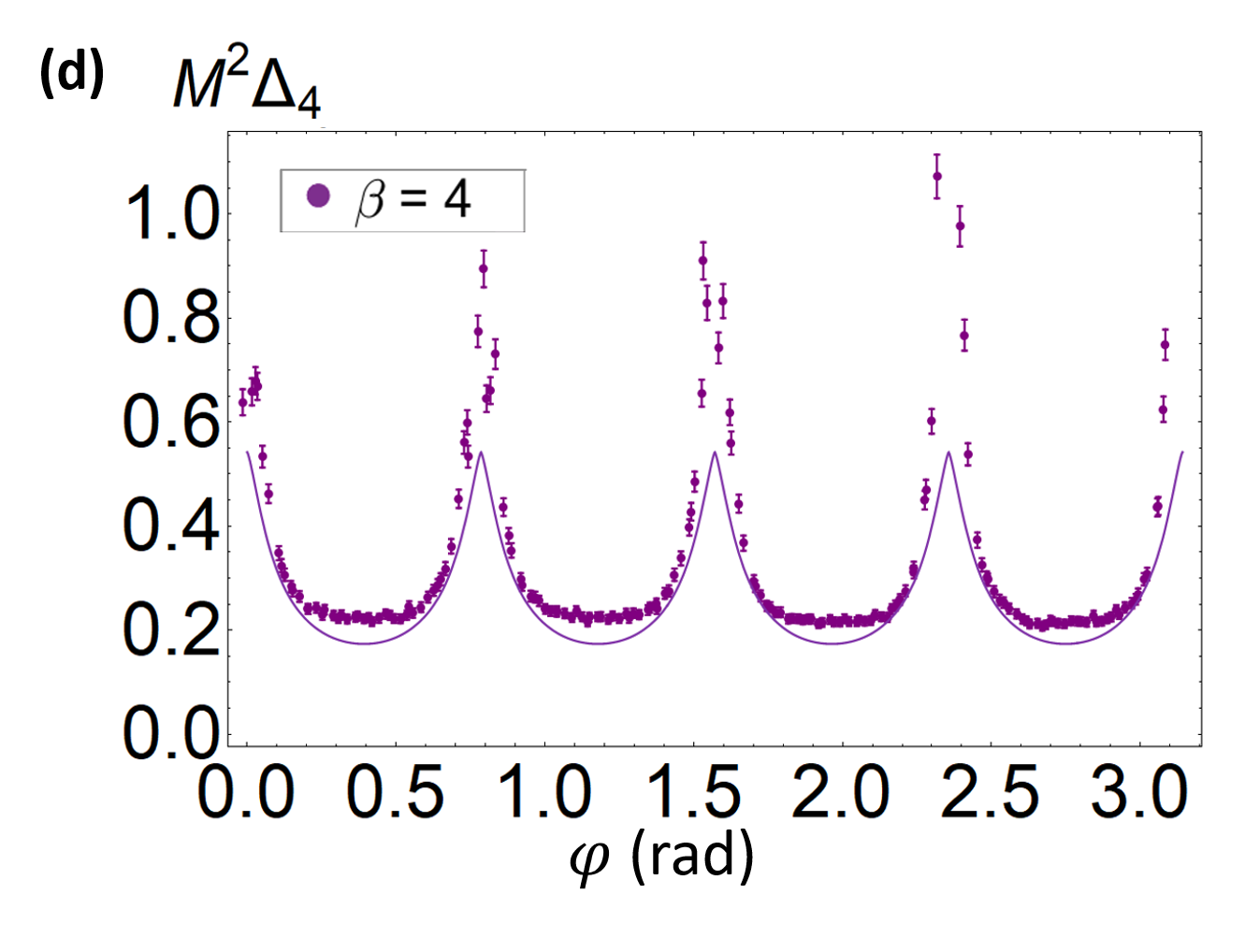}}
\caption{Absolute $\beta$ moments of the posterior distribution rescaled to the number of resources used in the experiment (M) as a function of the estimated phase value, fixing the value of the estimated visibility to $v_{\sf est} = 0.98$. The continuous line represents the generalized CRBs namely the inverse of the generalized FI to the power of $\beta$/$\alpha$ (see Eq. \ref{eq:gCRB}).}
\label{fig1}
\end{center}
\end{figure}

In Fig.~\ref{fig1} we report the estimated $\beta$ moments, for $v_{\sf est} = 0.98$, weighted with the number of resources, for every value of $\varphi$. We compare the estimation with its relative theoretical generalized CRB. In this case, the CRB for $\beta = 2$ is saturated almost everywhere. The most pronounced discrepancies occur for the highest values of $\varphi$ collected at the end of the experiment. There, the actual value of $v_0$ has a larger difference from $v_{\sf est}$, reflecting in the higher dispersion. However, even in this case, for some values of the phase the CRB is still achieved. This is due to the fact that the sensitivity to biases depends on the phase estimated, more precisely, as we have also noticed from the numerical simulations, the values of the phase that correspond to a maximum of the inverse of the FI (i.e. for $\varphi = 2k\frac{\pi}{8}$, with $k = 0,1,2,3...$) are more affected by small deviations from the real value and in fact show a more pronounced discrepancy from the CRB. On the contrary, phase measures corresponding to a minimum of the same function (i.e. for $\varphi = (2k+1)\frac{\pi}{8}$, with $k = 0,1,2,3...$) are more robust against noise and then less informative in the presence of biases. As for the bounds relative to the other moments of the distribution, the only one that is saturated, beyond $\beta = 2$, is the one linked to the third moment ($\beta = 3$). This can not be attributed to a different convergence of the moments to their relative expected values: previous work has demonstrated that all bound are saturated for at least $M\simeq 1000$~\cite{Review}, and in our experiment we have $M \simeq 10 000$.

\begin{figure}[H]
\begin{center}
{\includegraphics[width=.45\columnwidth]{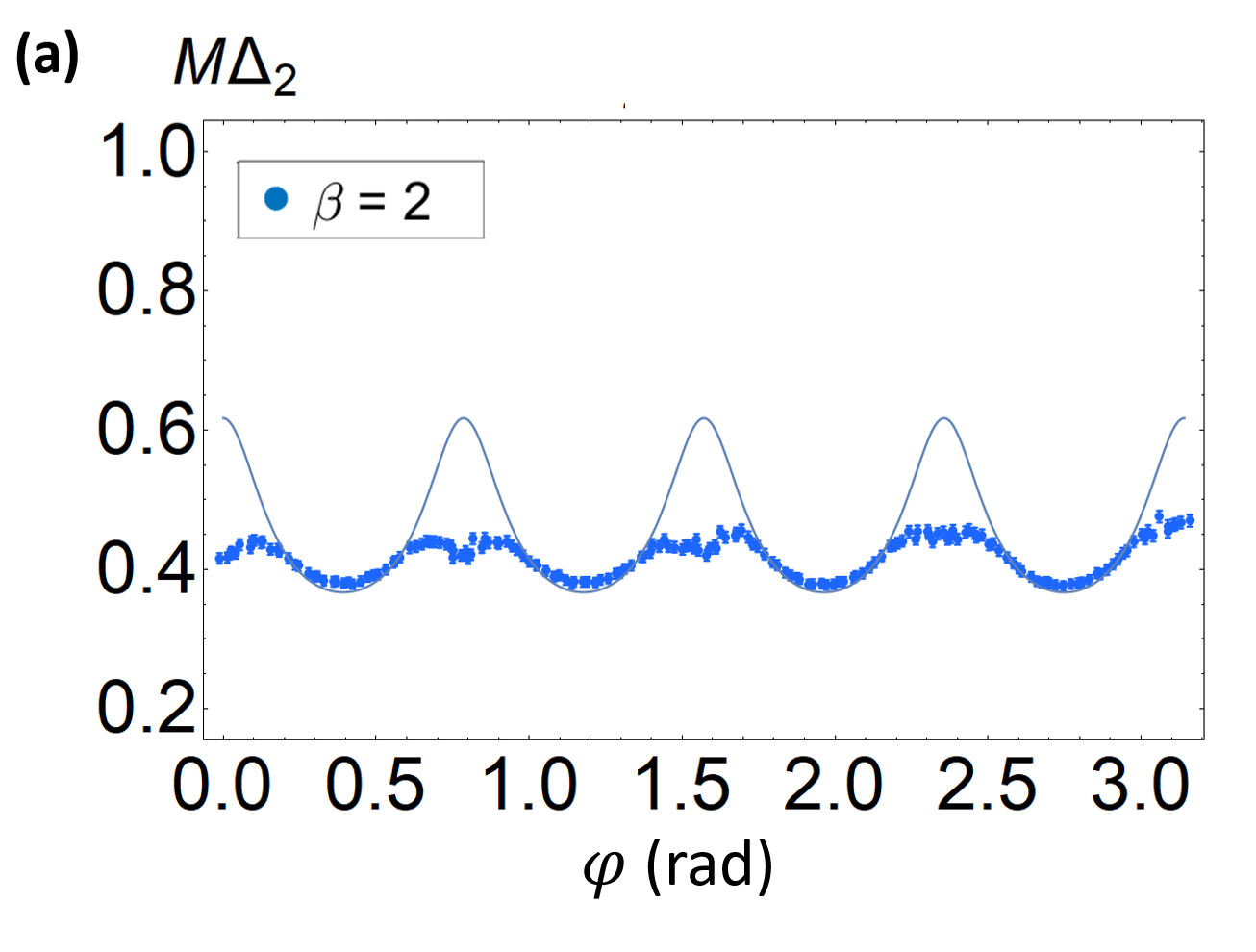}} \quad
{\includegraphics[width=.45\columnwidth]{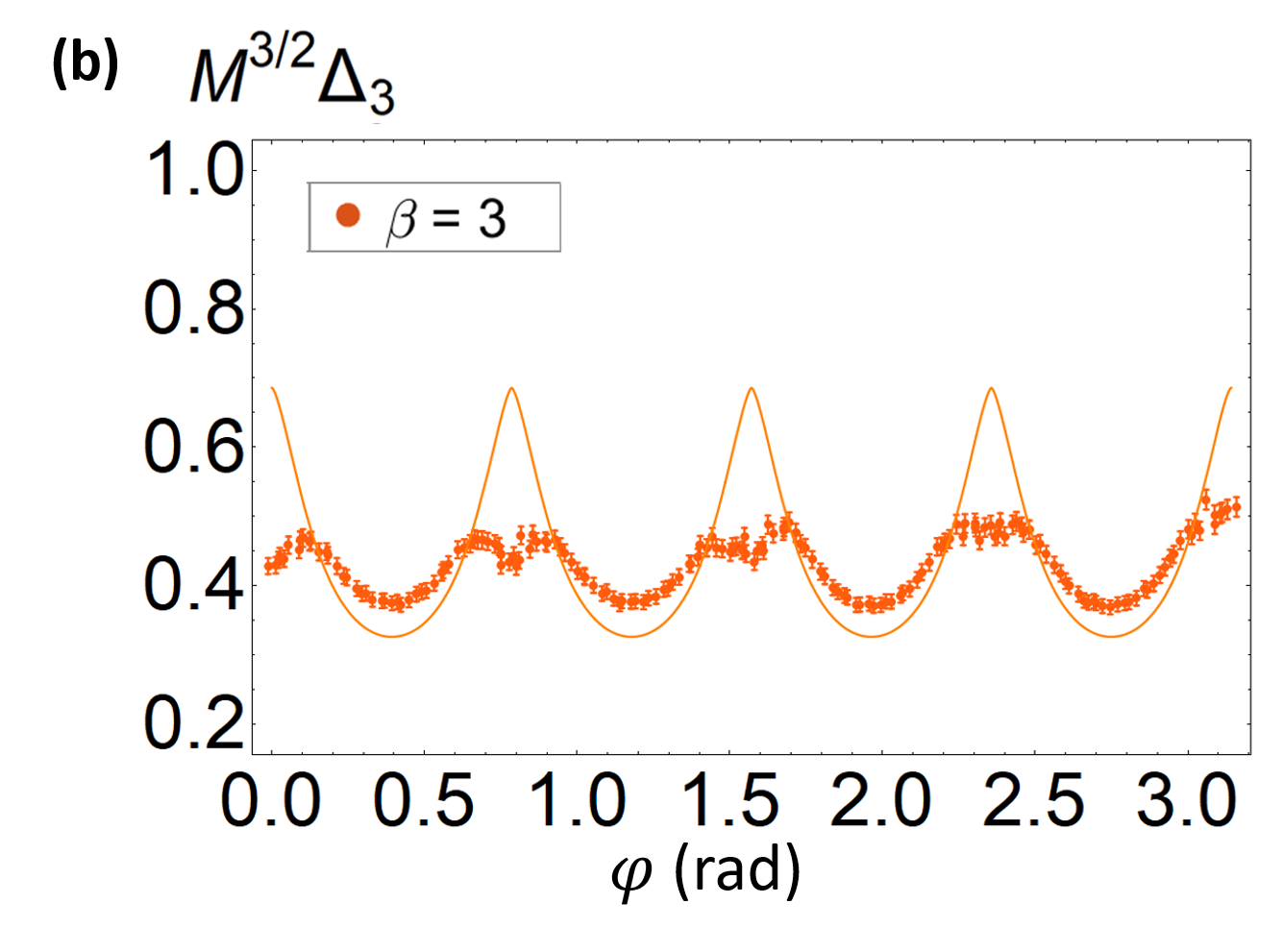}} \\
{\includegraphics[width=.45\columnwidth]{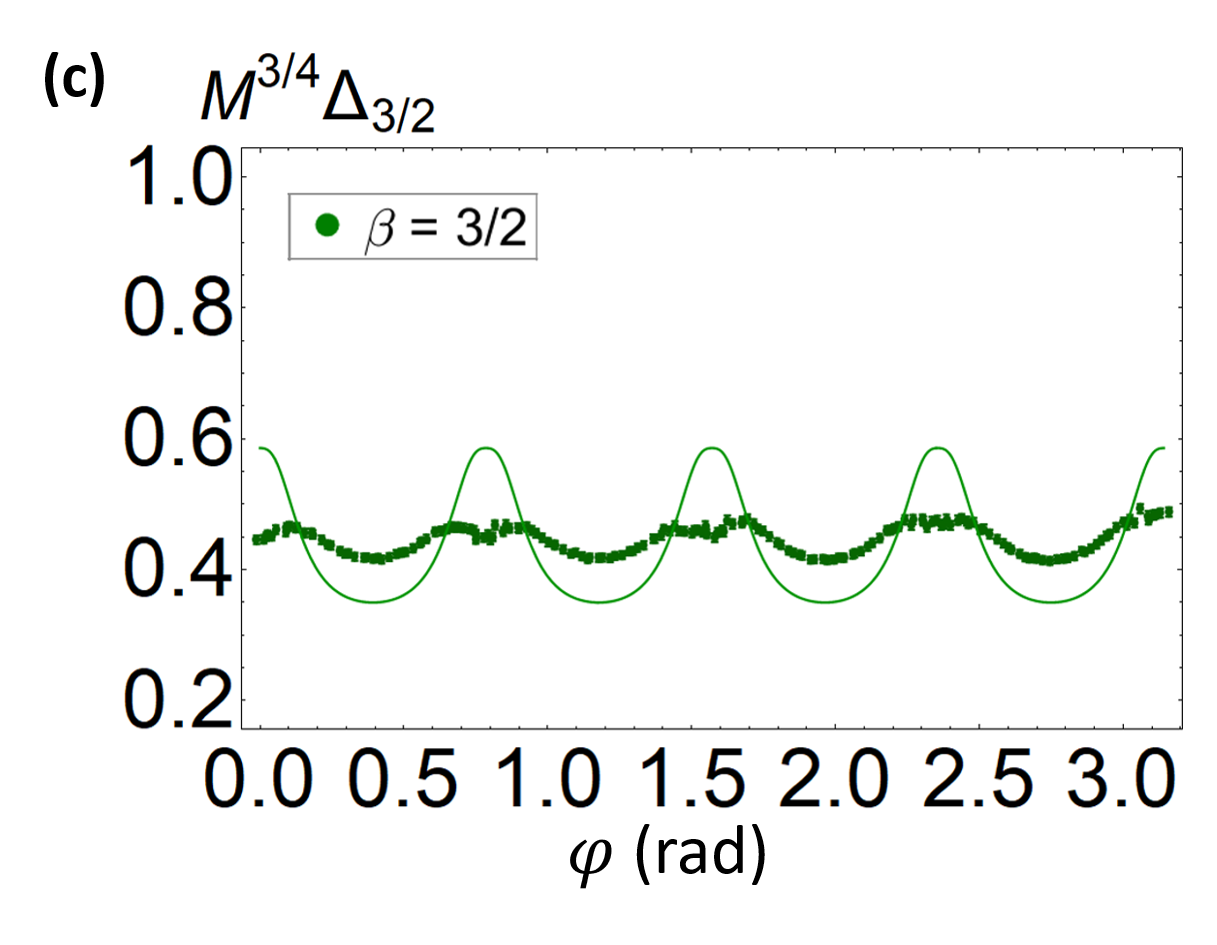}} \quad
{\includegraphics[width=.45\columnwidth]{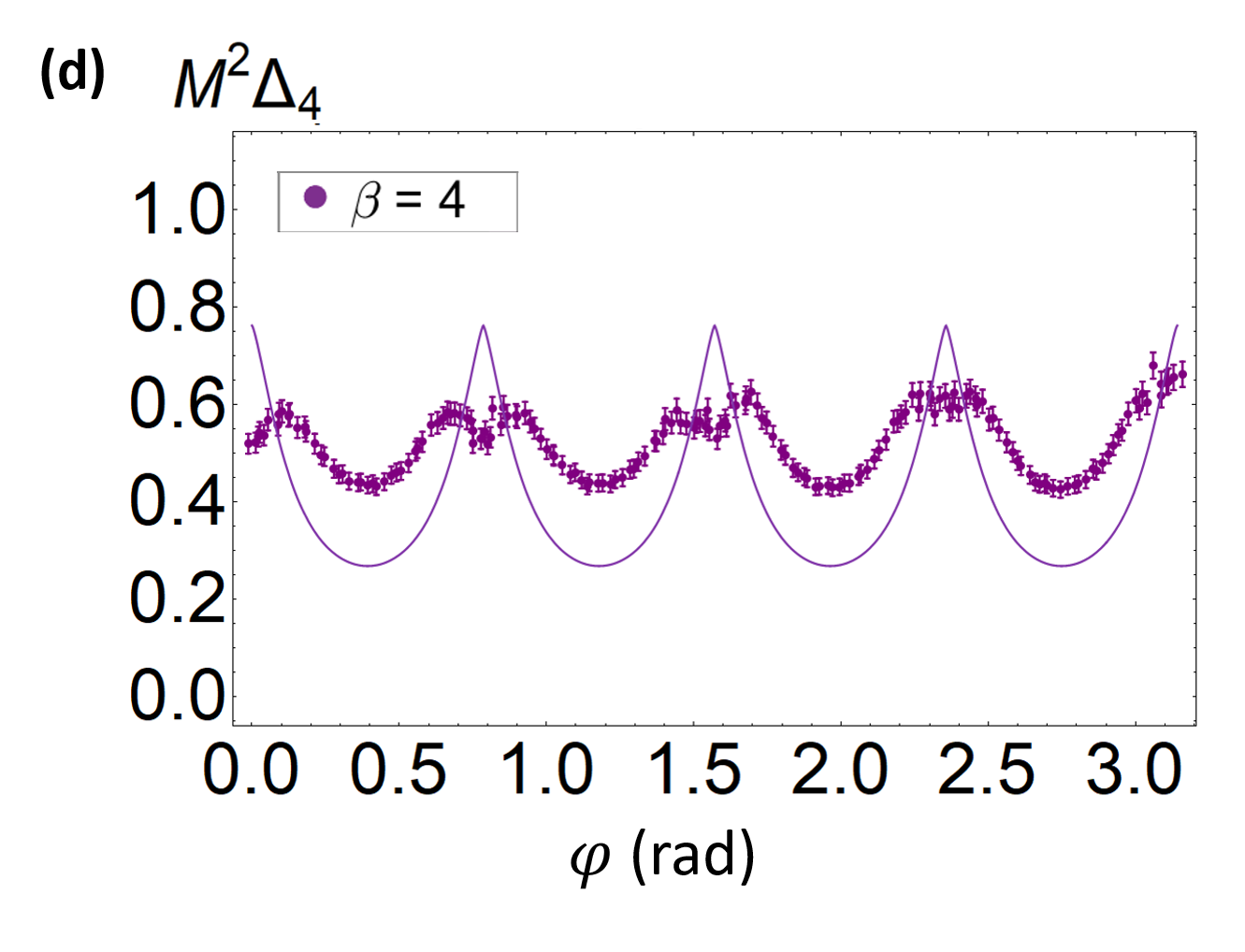}}
\caption{$M^{\beta/2}\Delta_{\beta}$ as a function of the estimated phase value, fixing the value of the estimated visibility to $v_{\sf est} = 0.90$. }
\label{fig2}
\end{center}
\end{figure}

In Fig.~\ref{fig2}, the same results are presented for a fixed visibility that is consistently lower than the actual one, i.e. $v_{\sf est} = 0.90$. There are instances in which the CRB with $\beta = 2$ is reached despite the inaccurate visibility, showing its limitations in revealing biases in the estimation procedure regardless the phase. In the same conditions it can be helpful to look at the bounds relative to the other $\beta$ and inspect if for some phase value they are more informative than the standard CRB.

\begin{figure}[H]
\includegraphics[width=\columnwidth]{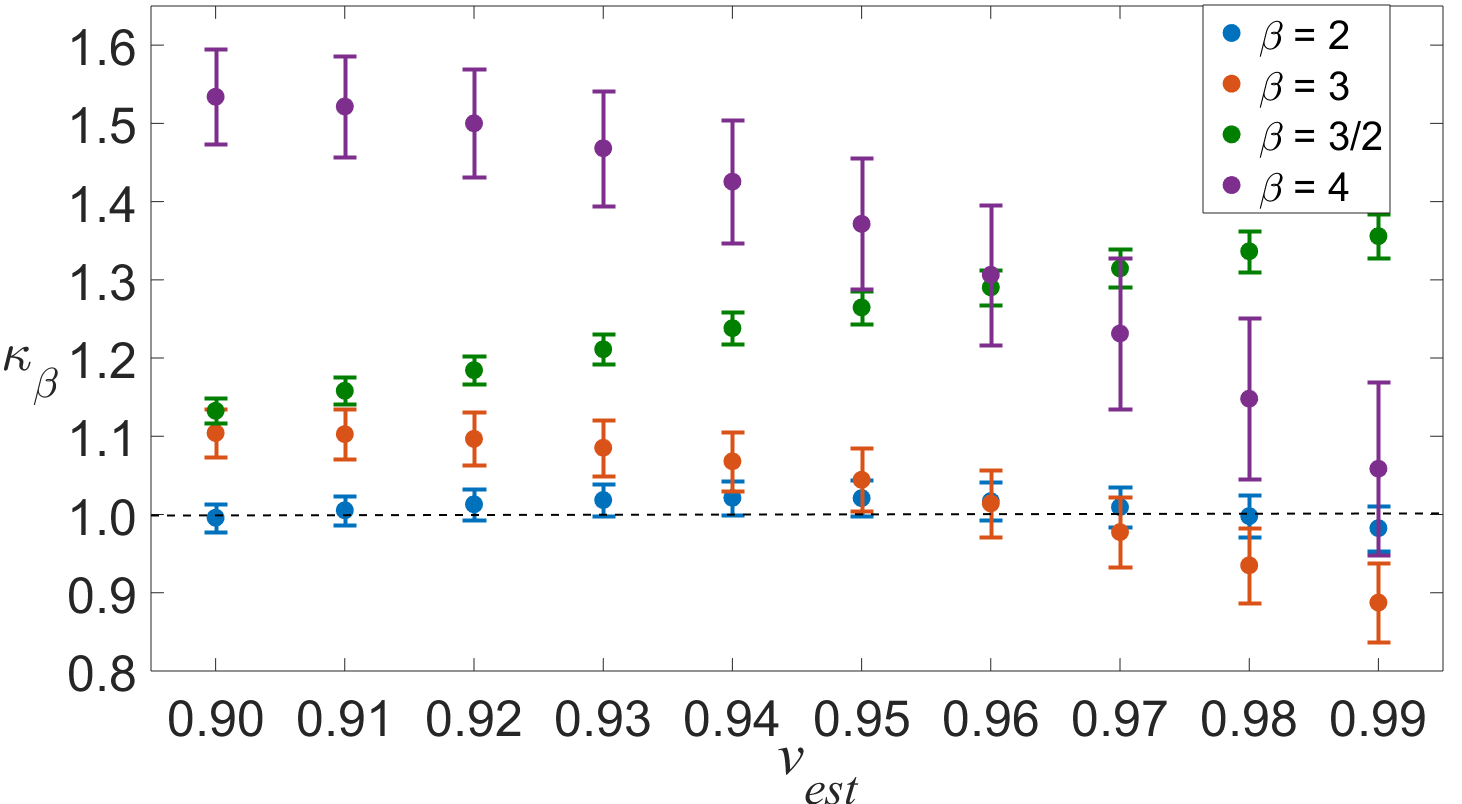}
\caption{$\kappa_\beta$ as a function of the pre-estimated value of the interferometer visibility $v_{\sf est}$, and for different values of the moment order, fixing the value of $\varphi = 2.8$ rad. The dashed line represents the generalized CRB.}
\label{fig3}
\end{figure}

In Fig.~\ref{fig3} we report the moments of the phase estimation distribution as a function of the value $v_{\sf est}$ imposed in the conditional probability. We plot the quantity
\begin{align}
\kappa_\beta = \frac{\Delta_\beta}{f_\alpha^{-\frac{\beta}{\alpha}}M^{-\frac{\beta}{2}}} \geq 1 \,,
\end{align}
whose lower bound equal to $1$ corresponds to the generalized CRB in Eq. (\ref{eq:gCRB}).

The estimation is now performed at a fixed value of $\varphi\simeq2.8$ rad when the visibility of the interference fringes has decreased to $v_{\sf true}\simeq0.96$. While the generalized CRB is never reached for the cases with $\beta = 3/2$ and $\beta=4$, the standard CRB is almost saturated for all the values of the visibility chosen, and in particular when $v_{\sf est}=v_{\sf true} \simeq 0.96$. It is interesting to see instead there is a region, corresponding to $v_{\sf est} > 0.97$, in which the bound for $\beta=3$ is violated while the bound relative to $\beta = 2 $ is fulfilled. Consistently with the numerical results presented in the previous section, this shows that the moment of order $\beta=3$ can give an indication on the presence of a possible bias not identifiable by looking only at the standard CRB. This happens for all the phase values except for those close to the minimal CRB. As previously illustrated by the numerical simulations, in that region it is harder to detect any violation of CRBs.

With these moments at hand, it is interesting to assess whether such bias affects the Gaussian shape of the probability distribution by comparing the measured moments to those expected $ \Delta_3^{\sf G} = 2\sqrt{\frac{2}{\pi}}(\Delta_2)^3$ and $ \Delta_4^{\sf G} = 3(\Delta_2)^4$. However, even when a value of $v_{\sf est}$ different to the actual value is employed in the data processing stage, the ratios $\Delta_3/\Delta_3^{\sf G}$ and $\Delta_4/\Delta_4^{\sf G}$ remain close to $1$. This is obtained for all cases in Fig.\ref{fig3}, demonstrating how the Gaussian approximation is unaffected and thus making this strategy not effective for detecting biases.

\section{Conclusion}

The importance of data post-processing in quantum parameter estimation can not be overemphasized. Finding the optimal estimator is as crucial as optimizing the experimental sequence and one should be able to detect inefficient possible biases in the model used for data analysis. While the focus is mostly on the CRB, conveying most of the information due to the Gaussian shape of the probability distribution, inspection of other moments may supply insight in this sense. This is however limited to orders which are close to saturate their bounds, such as $\beta=3$. In fact, we have shown that in some instances this moment results more sensitive to biases than the CRB. It is known that biases may appear due to a too small set of repetitions: this technique provides a possible real-time control for deciding when to stop the acquisition. Concerning different sources of bias, in particular those due to inaccurate modelling for data inversion, there exist no general prescriptions for quick fixes. In quantum polarimetry the adoption of multiparameter strategies has proven somewhat beneficial \cite{Roccia:18} with a consumption of extra-resources which can be kept modest \cite{Cimini19}.

Further perspectives of application of our findings can be found in adaptive estimation protocols \cite{Lumino}, which have been shown to provide convergence to the ultimate precision limits for limited number of probes.

Our work suggests that a more complete investigation even at a classical statistical level of phase estimation may provide relevant insights of the physics behind the estimation process. The power of this approach in any case is limited by the fact that in Bayesian estimation the final probability distribution always converges to a Gaussian. Further progress might leverage from alternative strategies to derive a distribution for the phases.
 
\begin{acknowledgments}
We thank A.~Smerzi for making us aware of the work~\cite{barankin1949}, and F.~Albarelli, M.G.A.~Paris and M.~Sbroscia for helpful discussion.
MGG acknowledges support from a Rita Levi-Montalcini fellowship of MIUR.
NS and FS acknowledge support from the Amaldi Research Center funded by the Ministero dell'Istruzione dell'Universit\`a e della Ricerca (Ministry of Education, University  and  Research) program Dipartimento di Eccellenza (CUP:B81I18001170001).
IG, NS and FS acknowledge Regione Lazio programme \emph{Progetti di Gruppi di ricerca legge Regionale} n. 13/2008 (SINFONIA project, prot. n. 85-2017-15200) via LazioInnova spa.

\end{acknowledgments}

\end{document}